\documentstyle[a4p,12pt,cite,psfig]{article}
\def\beq{\begin{equation}}  
\def\eeq{\end{equation}}
\def\bea{\begin{eqnarray}}
\def\eea{\end{eqnarray}}
\def\eq#1{{Eq.~(\ref{#1})}}

\begin{document}
\begin{titlepage}
\vspace{4mm}
\bigskip
\begin{flushright} 
  TAUP-2549-99\\

  January 25, 1999 \\
\end{flushright}

\vspace{7mm}

\bigskip 
\begin{center}
{\LARGE{\bf The dependence of the emission size}}\\
\vspace{3mm}  

{\LARGE{\bf on the hadron mass}}\\
\vspace{12mm}

{\bf Gideon Alexander}\footnote{\tt alex@lep1.tau.ac.il}{\bf ,} \
{\bf \ Iuliana Cohen}\footnote{\tt cohen@lep1.tau.ac.il}{\bf ,} \
{\bf \ Eugene Levin}\footnote{\tt leving@post.tau.ac.il}\\
\end{center}  
\vspace{2mm}

\centering{\it School of Physics and Astronomy}\\
\centering{\it Raymond and Beverly Sackler Faculty of Exact Sciences}\\
\centering{\it Tel-Aviv University, 69978 Tel-Aviv, Israel}\\

\vspace{20mm} 

\begin{abstract}
\vspace{5mm}  

The size of the emission volume of pions and kaons,
$r_{\pi \pi}$ and $r_{KK}$,  were
measured in the hadronic Z$^0$ decays
through the two-pion and two-kaon Bose-Einstein correlations 
near threshold. Recently
the emitter size  of the two identical baryons, $\Lambda \Lambda
(\bar \Lambda \bar \Lambda)$, was evaluated from the dependence
of the fraction of the spin S=1 state on the energy near threshold
where it is affected by the Pauli exclusion principle.
Here we show that the $r$ dependence on the particle masses, namely
the hierarchy  
\mbox{$r_{\pi \pi }\  > \ r_{K K}\ > \ r_{\Lambda \Lambda}$}  observed
in the Z$^0$ hadronic decays,
is well described in terms of the Heisenberg uncertainty principle.
A good description can also be obtained via
the virial theorem when applied to a general QCD potential.
Other available approaches are
also discussed.
\end{abstract}

\vspace{2.0cm}

\begin{center}
{\large Phys. Lett. B ~(in print)}
\end{center}
\end{titlepage}

\section{Introduction}
\label{intro}

It has been known for more than four decades
that the correlation function of two identical
particles allows us to learn about the size of the particle emission
region (the Hanbury Brown and Twiss effect \cite{HBT}).
The interference effect originated from Bose-Einstein
(Fermi-Dirac) statistics leads to an enhancement (reduction) of the
two-particle correlation  function which occurs  when two identical
particles are bosons (fermions) and have almost equal momenta.
The relation to the size of the emission region is often given by the
well known  
formula for the  correlation function of two identical particles with four
momenta $q_i \ ( i = 1, 2 )$ and $Q\ = \ \sqrt{-(q_1 - q_2)^2}$, namely
\beq \label{BOWLER}   
\sigma_{tot} \frac{d^2 \sigma_{12}}{d \sigma_1
\,d \sigma_2}\,\,\equiv\,\,C(Q)\,\,\,=\,\,1\,\,\pm\,\,|\,R(Q)\,|^2\,\,,
\eeq
where the sign $+\ (-)$ corresponds to bosons (fermions)
and $R(Q)$ is the
normalised Fourier transform of $\rho $, the source density
\cite{BOWLER}
\beq \label{SD}
R(Q)\,\,=\,\,\frac{\int dx\, \rho(x)\, e^{i \, (q_{1} - q_2) \cdot x}}
{\int dx\, \rho(x)\,}\,\,.
\eeq
This quantity is often parametrised in terms of a ``source radius" $r$,
related to  the size of the emission region,  and
a ``chaoticity parameter " $\lambda $, which measures the strength of the
effect. From Eq. (\ref{SD}) one derives that:

\beq \label{PAR}
|\,R(Q)\,|^2\,\,=\,\,\lambda\,e^{ - r^2 Q^2}\,\,,
\eeq

\noindent
assuming a spherical emission volume with a Gaussian density
distribution.
The study of the
Bose-Einstein Correlations (BEC)
of identical boson pairs, is carried out
by forming the ratio of the data correlation function
to the same correlation of data or Monte Carlo (MC) sample
having all the correlations, like those arising from resonances
apart from those due to BEC.
Since the original study of Goldhaber et al. \cite{goldhaber}
in 1959 and 1960, the BEC of pion pairs were studied in
a large variety of interactions over a wide range of energies
\cite{wolf}. In addition, studies of BEC of kaon pairs have also
been reported and values for their emitter dimension have been
evaluated. The study of this dimension has recently been extended
to pair of baryons, namely the
$\Lambda \Lambda (\bar \Lambda \bar \Lambda)$ systems using the Pauli
 exclusion principle.
From these experimental results an interesting hierarchy of the size of the
emission volumes is observed, namely \
$r_{\pi \pi }\  > \ r_{K K}\ > \ r_{\Lambda \Lambda}$.
This volume size dependence on the hadron mass may give a new insight
to the non-perturbative QCD and the hadronisation process.
In this work we investigate this
dependence using $\pi \pi$,\  $K K$ and $\Lambda \Lambda$ pair samples
present in the
multi-hadron Z$^0$ decays. In Section
\ref{sec2} we summarise the current knowledge of the BEC parameters
obtained from the studies of the hadronic Z$^0$ final states and
describe our procedure to extract representative $r_{\Lambda \Lambda}$ values. 
In Section \ref{sec3} we show that by using the
Heisenberg uncertainty principle one is able to describe
successfully the
$r$ dependence on the particle mass value.
In Section \ref{sec4} we
extract the mass dependence of the size of the interaction volume which
follows from the virial theorem by using
a general QCD potential. Finally the information on   
the interaction character that can be derived from the identical particle
correlations is summarised and discussed in Section 5.

\section{Extraction of the two-particle emitter size}
\label{sec2}
During the last decade values for
$r_{\pi \pi}$ have been reported by the ALEPH \cite{aleph_pipi},
DELPHI \cite{delphi_pipi} and OPAL \cite{opal_pipi}
collaborations from measurements in the hadronic Z$^0$ decay
events. These $r$ values are summarised in Table \ref{tab_2pions}.
To notice is the fact that the $r$ values are rather sensitive to the
choice of the reference sample which
was either the data $\pi^+ \pi^-$ system or a sample of
two-pions originating from two different events, the so called
mixed event sample. The ALEPH and DELPHI collaborations
have therefore opted to quote
their final result as the average of the
values obtained from these two methods. OPAL chose as the best $r$ value
the one obtained with a $\pi^+ \pi^-$ data reference sample.
The result obtained with a MC generated reference sample was used for the
estimation of the systematic error. To be on equal footing
with the ALEPH and DELPHI collaboration, we also present in
Table 1 the average OPAL value obtained from their two methods.
Finally we also list in the last row of Table \ref{tab_2pions} the
over-all averaged $r_{\pi \pi}$ value
of $0.74 \pm 0.01 \pm 0.14$ fm where the first error and
the second one are respectively the statistical and systematic uncertainties.
\renewcommand{\arraystretch}{1.6}
\begin{table}[thbp]
\begin{center}
\begin{tabular}{l c c c}
\hline
        &\multicolumn{3}{c}{ { \large  $r_{\pi\pi}$} (fm)  } \cr
\cline{2-4}
   Experiment  &  Method I & Method II  & I \& II Combined\cr
\hline \hline
ALEPH \cite{aleph_pipi} & $0.80 \pm 0.04 $ & $0.50 \pm 0.02$ & $0.65 \pm 0.04 \pm 0.16$\cr

DELPHI \cite{delphi_pipi} &$0.82 \pm 0.03 $ & $0.42 \pm 0.04$ & $0.62 \pm 0.04 \pm 0.18$\cr

OPAL \cite{opal_pipi}   &$0.96 \pm 0.01 $ & $0.79 \pm 0.02$ & $0.88 \pm 0.02 \pm 0.09$\cr

\hline \hline
LEP Average    & $0.90 \pm 0.01$ & $ 0.62
 \pm 0.01 $ & {$ \bf 0.74 \pm 0.01 \pm 0.14$}\cr

\hline 
\end{tabular}
\end{center}
\caption{The two-pion emitter size obtained from BEC studies using the
hadronic Z$^0$ decays carried out with two choices for the   
reference sample. In method I the reference
sample was the data $\pi^+ \pi^-$ system. In method II the reference
sample was created by the event mixing technique or Monte Carlo
generated 
events. In the last row the weighted average values are presented.   
The first and
the second errors are respectively the statistical and systematic
uncertainties. The systematic error is taken as half the difference
between the values of the two methods.}
\label{tab_2pions}
\end{table}
\renewcommand{\arraystretch}{1.0}

\noindent
Several BEC studies have been reported for the pairs of kaons. The
BEC parameters obtained for the $K^0_S K^0_S$ system do suffer from
the fact that part of the observed low $Q$
enhancement may originate from the decay of the $f_0(980)
\to K^0 \overline{K}^0$. On the other hand the BEC of the $K^{\pm} K^{\pm}$
pairs are free of this problem. In the following whenever we
refer to a kaon pair we mean the  $K^{\pm} K^{\pm}$ systems.
Using the Z$^0$ hadronic decays,
DELPHI  \cite{delphi_kk} has measured the BEC of these
$K^{\pm} K^{\pm}$ pairs and obtained
results for the emitter dimension $r_{KK}$
and the strength $\lambda_{KK}$:
$$r_{K K} = 0.48 \pm 0.04(stat.) \pm 0.07(syst.) \ {\rm{fm}}
\ \ {\rm{and}} \ \ \lambda_{K K} = 0.82 \pm 0.11(stat.) \pm
0.25(syst.)$$ 

\noindent
A method for the measurement of the emitter dimension
of two identical baryon systems in multi-hadron final state
has recently been proposed
by G. Alexander and H.J. Lipkin
\cite{alexlipkin}. This method, which is applicable
to pairs of identical spin 1/2 baryons which decay weakly, uses
the Fermi-Dirac statistics properties which suppress the over-all spin
S=1 state of the identical di-baryon system near
their threshold (Pauli exclusion principle). \\
\indent
To apply e.g. this method to $\Lambda \Lambda$ or
$\Lambda \bar \Lambda$ pairs, the hyperon-pair
is first transformed to their common centre of mass (CM) system and then each
$\Lambda$ decay proton
is transformed to its parent hyperon CM system.
The distribution of the cosine angle between these two protons, here 
denoted by $y^{\ast}$, depends on the fraction of the S=1 (or S=0) in the data.
Specifically the following $y^{\ast}$ distributions are expected
for pure S=0 and S=1 states of the $\Lambda$ pair
$$dN/dy^{\ast}|_{_{S = 0}} \ = 1 + (-1)^{B/2}
\cdot \alpha_{\Lambda}^2 \cdot y^{\ast} \ \ \ \ {\rm{and}} \ \ \ \
dN/dy^{\ast}|_{_{S = 1}} \ = 1 - (-1)^{B/2}
\cdot \alpha_{\Lambda}^2 \cdot y^{\ast}/3 \ ,$$

\noindent
where $\alpha_{\Lambda}$=0.642$\pm$0.013 \cite{pdg98}
is the $\Lambda \to p \pi^-$ decay asymmetry parameter arising from parity violation
and B is the baryon number of the di-hyperon system.
These distributions are independent of the orbital angular
momentum and are valid as long the $\Lambda$'s are non-relativistic
in their CM system. Therefore in this method the spin composition
analyses are  
restricted to relatively small $Q$ values\footnote{Assuming that no
exotic $\Lambda \Lambda$ resonances exist in this Q region.}.
Defining $\varepsilon$ as the fraction of the S=1 contribution
to the di-hyperon pairs, it is equal to
 \[ \varepsilon =\frac{(S=1)}{(S=1)+(S=0)}\ , \]
so that the following function:

$$dN /dy^{\ast}=f_{BG} \ + \
(1-f_{BG})\cdot {\{}(1-{\bf \varepsilon})\cdot dN / dy^{\ast}|_{_{S = 0}}
+\varepsilon \cdot dN / dy^{\ast}|_{_{S=1}}{\}}\ ,$$

\noindent
can be fitted to the data at every $Q$ values.
Here $f_{BG}$ is the background fraction in the data. As in  BEC 
analyses the possible, if at all, final states interactions are neglected.\\  
\indent  The fraction of the S=1 spin content,  
$\varepsilon$,  of the
$\Lambda \Lambda (\bar \Lambda \bar \Lambda)$
system  has been measured as a function of $Q$
in $\sim 4 \times 10^6$  hadronic Z$^0$ decays per experiment at LEP
by the OPAL \cite{Lam1}, ALEPH \cite{Lam3}
and DELPHI \cite{Lam2} collaborations. A decrease in $\varepsilon$
has been observed as $Q$ approached zero in contrast to the results
obtained for the $\Lambda \bar \Lambda$ where $\varepsilon$
was found to be equal to 0.75 down to very low $Q$ values.
The $r_{\Lambda \Lambda}$ values deduced by
the OPAL and DELPHI collaborations are : \\

\begin{flushleft}
\vspace{-1mm}

\hspace{4.5cm} OPAL \cite{Lam1} \ \ \ \ \ \ \
$r_{\Lambda\Lambda}  = 0.19 ^{\ + 0.37}_{\ -0.07} \ \pm \ 0.02 ~~{\mathrm fm}\ ,$
\vspace{3mm}

\hspace{4.5cm} DELPHI \cite{Lam2} ~~~
$r_{\Lambda\Lambda}
 = 0.11 ^{\ + 0.05}_{\ - 0.03} \ \pm \ 0.01 ~~{\mathrm fm}\ .$\\
\end{flushleft}

\noindent
The ALEPH results for the $\Lambda\Lambda$ spin composition
 are similar to those of OPAL and DELPHI
but no attempt has been made to extract from them an
\mbox{$r_{\Lambda \Lambda}$ value.} \\
\begin{figure}
\centerline{\hbox{
\psfig{figure=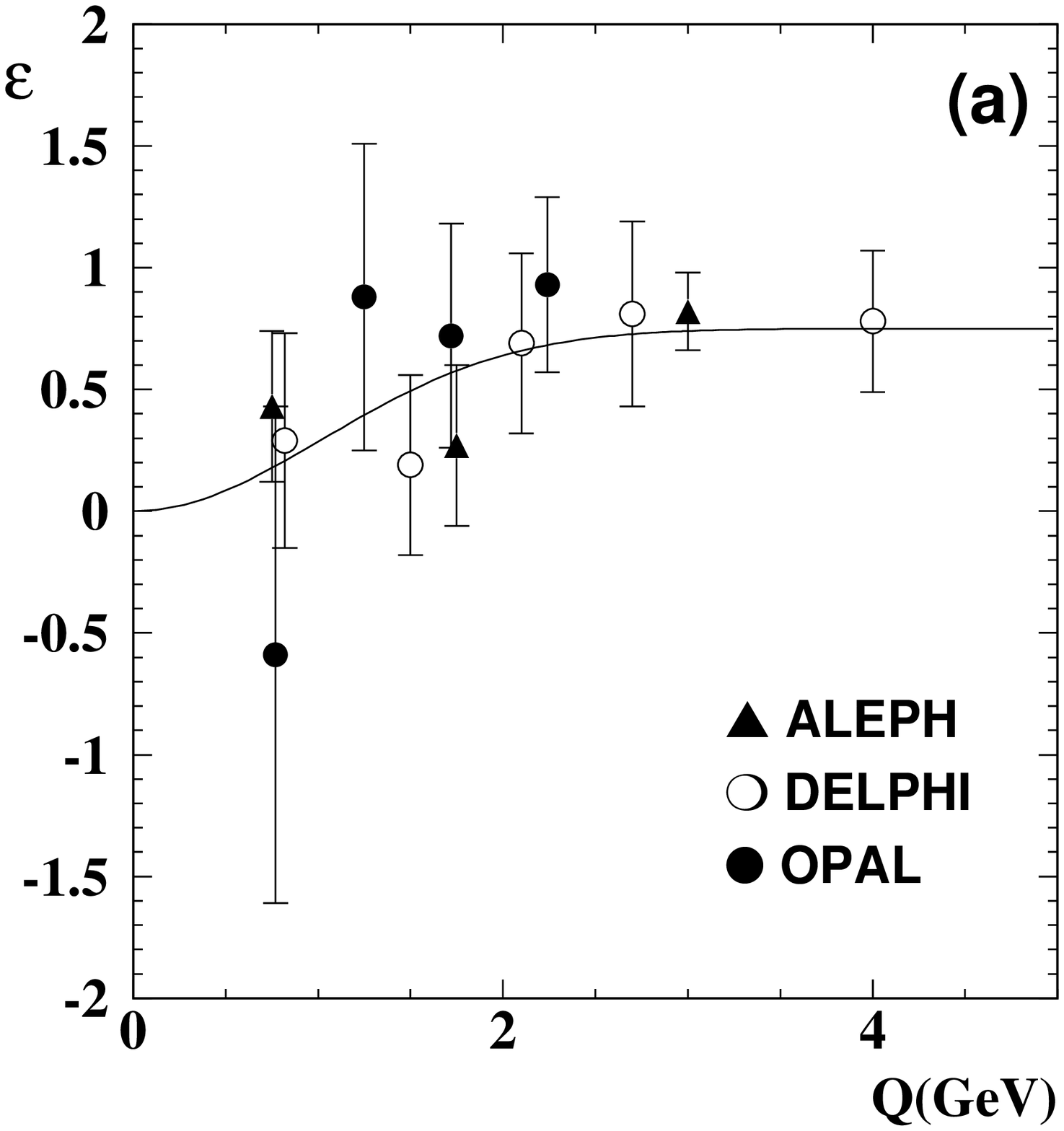,width= 8cm,height=8cm}
\psfig{figure=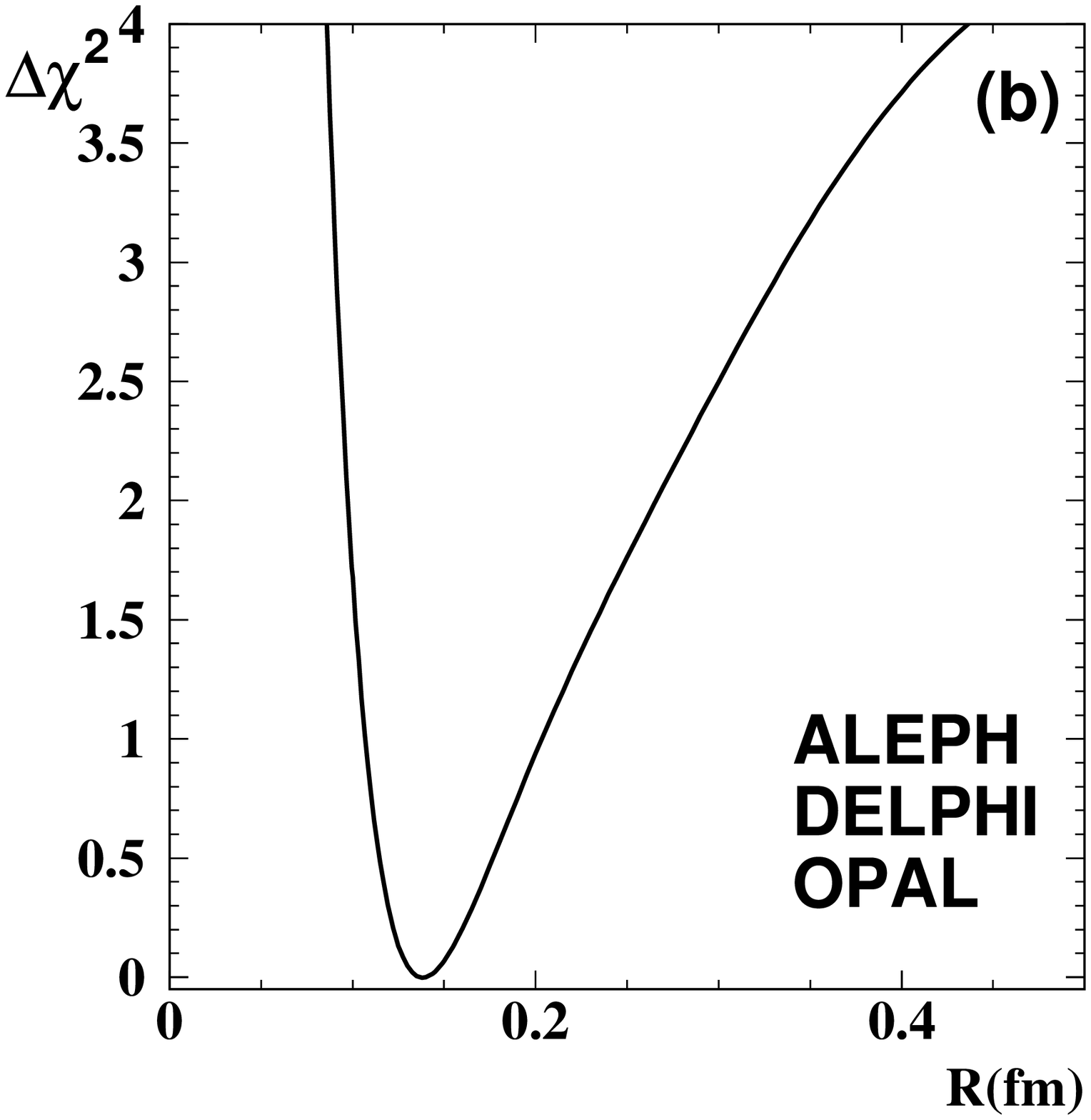,width= 8cm,height=8cm}
}}
\caption{(a)
The S = 1 fraction, $\varepsilon$, of the
$\Lambda\Lambda(\bar \Lambda \bar \Lambda)$ pairs
measured as a function of $Q$ by the ALEPH \cite{Lam3}, DELPHI  \cite{Lam2}
and OPAL  \cite{Lam1} 
collaborations. The solid line represents the fit results of
Eq. (\ref{pauli}) to the data points.
(b) The  $\Delta\chi^2 = \chi^2 - \chi^2_{min}$ dependence on
$r_{\Lambda\Lambda}$ as obtained from the fit to the combined LEP data.}
\label{fig_lamlam}
\end{figure}
\indent
The measured  $\varepsilon(Q)$ values of the three experiments
are compiled in Fig. \ref{fig_lamlam}a where the error bars are
strongly dominated by the statistical uncertainties.
The solid line in the same figure is the outcome of our
unbinned maximum likelihood fit of the function:
\begin{equation}
\varepsilon(Q) \ = \ 0.75
    [1\ - \ e^{-r^2_{\Lambda\Lambda}Q^2}]\ ,
\label{pauli}
\end{equation}
to the data points plotted
in the Fig 1a. The factor
0.75 appearing in Eq.~(\ref{pauli}) represents the statistical spin mixture
which is proportional to 2S + 1.
From this over-all
fit, with a $\chi^2/d.o.f.=0.4$, the dimension of the $\Lambda \Lambda$ emitter and its
errors are found to be
\[
r_{\Lambda\Lambda} \ = \ 0.14 ^{\ + 0.07}_{\ - 0.03} ~~{\mathrm
fm}\ . \]
\noindent
The $\Delta\chi^2 = \chi^2 -\chi^2_{min}$ behaviour of this fit is shown in
Fig. \ref{fig_lamlam}b.
The ALEPH collaboration has also used
an alternative method to measure  $r_{\Lambda\Lambda}$. In that method
one constructs, similar to the BEC studies, a correlation function of
the type
\[ C_2^{\Lambda \Lambda}(Q) = N^{\Lambda \Lambda}_{data}(Q)
/  N^{\Lambda \Lambda}_{ref}(Q)\ , \]
where the numerator is the data distribution and the denominator
is the distribution of a reference sample void of the Fermi-Dirac
statistics. This $C_2^{\Lambda \Lambda}(Q)$ correlation is expected to 
decrease at low $Q$ values due to the onset of the Pauli exclusion principle.
From this analysis ALEPH evaluates  $r_{\Lambda\Lambda}$ to be
$(0.09 - 0.10) \pm 0.02$ fm in perfect agreement with the values
obtained from the spin composition analyses.

\begin{figure}[t]
\centerline{
\psfig{file=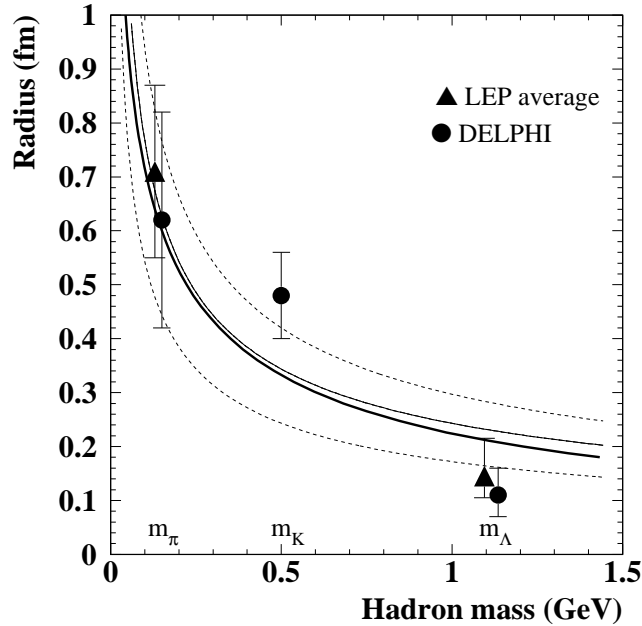,height=9.2cm,width=9.2cm}
}
\caption{The emitter radius $r$ as a function of the hadron
mass $m$. The  triangles  are the DELPHI results and the
circles are the averages of the measured values  at LEP.
For clarity the points are plotted at slightly displaced  mass values.
The error bars correspond to the statistical and systematic
errors added in quadrature. The thin solid
line represents the expectation from the Heisenberg uncertainty  
relations
setting $\Delta t = 10^{-24}$ seconds. The upper and the
lower
dashed lines correspond to $\Delta t = 1.5\times 10^{-24}$
and $0.5\times 10^{-24}$ seconds respectively. The thick solid line
represents the dependence of $r$ on $m$ as expected from a QCD 
potential given by Eq. (\ref{VT5}).} 
\label{main}
\end{figure}

\noindent
In Fig. \ref{main} we present the experimental results for $r$ obtained
in the  analyses of the hadronic Z$^0$ decays. 
The  DELPHI values, which is the only LEP experiment that measured ~$r_{KK}$,
are shown by circles.
The LEP averages of the measured ~$r_{\pi \pi}$               
and $r_{\Lambda \Lambda}$ values are also shown by triangle symbols
 and are seen to be in good agreement with DELPHI ones.
The large error associated with $r_{\pi \pi}$ reflects the systematic
uncertainty due to the
strong dependence of the experimental results on the choice
of the reference sample. In contrast to this situation, a relative
small systematic error is associated with $r_{\Lambda \Lambda}$
since in this case the spin analyses do not rely on a reference sample.
In spite of the relative large errors a general non-trivial
trend can be observed
in the $r$ dependence on the particle mass, namely that:
\[ r_{\pi \pi} \ > \ r_{K K}\  >\  r_{\Lambda \Lambda}\ .\]

\section{The emission volume size and the uncertainty relations}
\label{sec3}

The maximum of the BEC enhancement of two identical
bosons of mass $m$
occurs when $Q \rightarrow 0$ which also means that
the three vector momentum difference of the bosons approaches zero.
This motivated the interest to link the BEC effect to the uncertainty 
principle \cite{heinz}.
From the Heisenberg uncertainty relations we  have that

\begin{equation}
\Delta p\,\Delta r \,\, = \,\,2\,\mu\,\emph{v}\,r\ = \ m\,\emph{v}\,r\
= \  \hbar\,c \ ,
\label{heis1}
\end{equation}

\noindent
where $m$ and $\emph{v}$ are the hadron mass and its velocity.
Here $\mu$ is
the reduced mass of the di-hadron system and $r$ is the distance   
between them.
In Eq. (\ref{heis1}) the momentum $\Delta p$ is  measured in GeV,
$\Delta r \ \equiv \ r$ is given in fermi units
and   $\hbar c \ = \ 0.197$ GeV fm.
From this, one obtains

\begin{equation}
r\ = \ \frac{\hbar c}{m \emph{v}}\ = \ \frac{\hbar c}{p}\ .
\label{r2}
\end{equation}

\noindent
 Simultaneously we also use the uncertainty
relation expressed in terms of time and energy

\begin{equation}
\Delta E \Delta t \,\, =\,\,  \frac{ p^2}{m}\, \Delta t \,\, =\,\,
\hbar \ ,
\label{heis2}
\end{equation}

where the energy is given in GeV and $\Delta t$ in seconds.
Thus we have

\begin{equation}
p^2 = \hbar\,m/\Delta t \ \ \ \ {\rm{so \ that}} \ \ \ \
p  \,\, = \,\, \sqrt{\hbar\,\,m/\Delta t}\ .
\label{eqv}
\end{equation}

\noindent
Inserting this expression for $p$ in Eq. (\ref{r2})
we finally obtain

\begin{equation}
r(m) \ = \ \frac{\hbar c/ \sqrt{\hbar /\Delta t}}{\sqrt{m}}\ = \
\frac{c \sqrt{\hbar \Delta t}}{\sqrt{m}}\ .
\label{final}
\end{equation}
In the following we take $\Delta t = 10^{-24}$
seconds as a representative time
scale of the strong interaction which is assumed to be independent of the
hadron identity and its mass.
Thus $r(m) \ = \ A/\sqrt{m}$ \ with
$A$\ = \ 0.243 fm GeV$^{1/2}$. 

\noindent
In Fig. \ref{main} we show by the thin solid line
the $r$ dependence
on the mass as calculated
by Eq. (\ref{final}) setting $\Delta t = 10^{-24}$ seconds.
The sensitivity of
$r(m)$ on the value of $\Delta t$ is illustrated by the dashed lines in
the figure. The upper and the lower dashed lines correspond
respectively to the
values $\Delta t = 1.5\times 10^{-24}$ and  $0.5\times
10^{-24}$ seconds. A fit of Eq. (\ref{final}) to the
three data points plotted in Fig. \ref{main} yields
$\Delta t \ = \ (1.2 \pm 0.3)\times 10^{-24}$ seconds with
a $\chi^2 /d.o.f.$ = 4.2 which corresponds to a probability of \
$\simeq$12\%.
As seen from Fig. \ref{main}, the dependence of $r$ on $m$
as given in Eq. (\ref{final})   
describes, within errors, very well the mass dependence of the
measured emitter size
and offers a simple explanation for the measured
hierarchy $r_{\pi \pi} \ >\  r_{KK}\  > \ r_{\Lambda \Lambda}$.\\

\section{The emission volume size and the virial theorem}  
\label{sec4}

The previous section shows that the observed hierarchy $r_{\pi \pi} \ >\
r_{KK}\  > \ r_{\Lambda \Lambda}$ has a natural and a rather general
explanation. Here we would like to demonstrate that such a dependence
could shed  light on the character of the ``soft" interaction or, in other
words, on the non-perturbative QCD which is responsible for the 
interaction at long  distances.  Let us continue to use the semi-classical 
approximation as in section \ref{sec3}, which means that the angular
momentum
\beq \label{VT1}
\ell \,\,=\,\,|
\,\vec{p}_1\,\,-\,\,\vec{p}_2\,|\,b_t\,\,=\,\,2
\,p\,b_t\,\,\approx\,\,\hbar\,c\,\,,
\eeq
where $b_t$ is  the impact parameter. To estimate the value of $p$ we 
use the virial theorem \cite{VT} which yields a general connection
between the average values of the kinetic and the potential energies,
namely
\beq \label{VT2}
2\,\langle \,T_{t}\,\rangle \,\,=\,\,\langle \,\,\vec{b}_t
\,\cdot\,\vec{\nabla}_t\,V(r)\,\,\rangle \,\,,
\eeq
where $t$ denotes the transverse
direction. Since the motion in the transverse direction is always
finite, we can safely use the virial theorem.
Substituting in \eq{VT2} the kinetic
energy by its relation to the momentum, $T_t\,\,=\,\,p^2_t/ m$, we obtain
\beq \label{VT3}
\langle \ p^2_t\,\rangle \,\,= \,\,m\,\langle \,\,\vec{r}_t
\,\cdot\,\vec{\nabla}_t\,V(r)\,\,\rangle\,\,,
\eeq
where we denote by $r$ the distance between two particles with $r_t $
being equal to $2 b_t$.  \\
\indent From \eq{VT3} one can see that the relation between
$\langle p^2\,\rangle$ and $r$
depends crucially on the $r$ dependence of the potential energy $V(r)$.
Here we take  $V(r)$ to be independent 
of the interacting particles (quarks) mass. 
This assumption is also in accordance with the Local Parton 
Hadron Duality \cite{DUAL} concept, which states
that one can consider the production of hadron as interaction of partons, 
i.e. gluons and quarks, in spite of the unknown mechanism of hadronisation. 
This approach is supported experimentally by the single and 
double inclusive
hadron production in $e^{+ }\,e^{-}$ annihilation \cite{DUAL}. 
Since the interaction
between partons is independent of the produced hadron mass it follows that 
$\partial V(r)/\partial m = 0$.
Based on this assumption and using
\eq{VT1} and \eq{VT2},
we derive an equation for the typical size of the emission volume, namely
\beq \label{VT4}
r^2\,\,\langle\,\,\vec{r}
\,\cdot\,\vec{\nabla}\,V(r)\,\rangle \,\,\,\approx\,\,\, \frac{(\,\hbar
\,c\,)^2}{m}\ . 
\eeq
To illustrate the kind of information that can be extracted
from \eq{VT4}, we
evaluate the typical size of the particle emission region from
two very different potential energy $V(r)$ forms:

\begin{enumerate}
\item\,\,\, $V(r)\,\,=\,\,N_c\alpha_s \hbar c / (\pi \,r)$\
which models the
QCD induced interaction at short distances.
In this case, from \eq{VT4} one obtains $r
\,\,\propto\,\,\hbar\,c / (\bar \alpha_S \,\,m)$.

\item\,\,\,$V(r)\,\,=\,\,\kappa\,r$ which corresponds to the confinement
potential.
For this potential \eq{VT4} leads to $r \,\,\propto \,\,1/m^{\frac{1}{3}}$.

\end{enumerate}

\begin{figure}[hbt]
\centerline{
\psfig{file=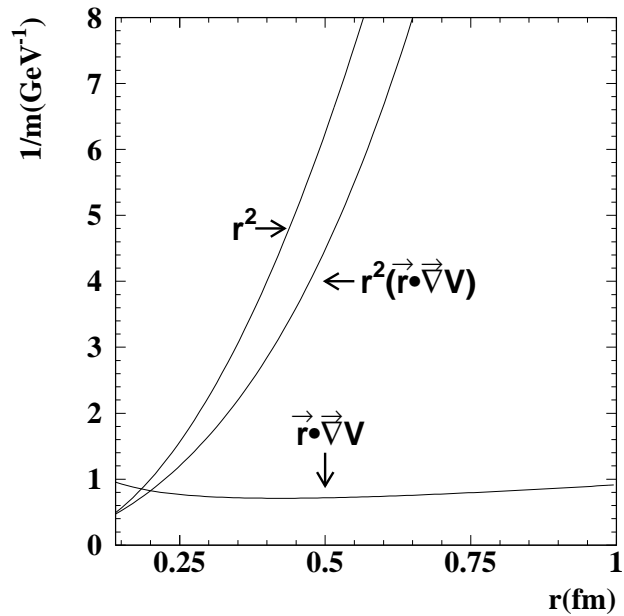,height=9cm,width=9cm}
}
\caption{The behaviour of three factors,
$\vec{r} \cdot \vec{\nabla}V$, $r^2 (\vec{r} \cdot \vec{\nabla})$
and $r^2$ which contribute
to the dependence of $1/m$ on the radius $r$ when using the
potential $V(r)$ defined in Eq. (\ref{VT5}).
}
\end{figure}

\noindent
It is interesting to note that the experimental data plotted in Fig. 2
are well described by \eq{VT4} when applied to the general
potential form   
\beq \label{VT5}
V(r)\,\,\,=\,\,\,\kappa\,\,r\,\,\,-\,\,\,\frac{4}{3}\,\frac{\alpha_S
\hbar c}{r}\,\,,
\eeq
which is widely used to derive the wave functions and decay constants
of hadrons \cite{RCP}. This can be seen by comparing in Fig. 2
the thick solid line with
the experimental data on the radii
derived from the identical particles
correlations. This line was obtained by setting \eq{VT4} to an
equality and 
evaluating \eq{VT5} with the $V(r)$ parameter values
of $\kappa = 0.14$ \ GeV$^2$ = 0.70 GeV/fm\  and
$\alpha_s \,\,= 2 \pi /9 \,\ln(\delta + \gamma/r )$ with $\delta = 2 $ and
$ \gamma = 1.87$ \ GeV$^{-1} = 0.37$ fm obtained from the hadron wave functions and 
decay constants \cite{RCP}. \\
\indent
Here it should be stressed that the simple \
$r\,\propto\,1/\sqrt{m}$ \ dependence obtained
from the uncertainty principle
does not contradict the virial theorem estimate.
The fact that for the QCD potential defined by Eq. (\ref{VT5}), the factor
 \mbox {$\langle \vec{r}\,\cdot\,\vec{\nabla}_t\,V(r)\,\rangle$} is essentially
constant (see Fig. 3) over the $r$ range 0.14 to 1.0 fm, means again
that \ $r~\propto~1/\sqrt{m}$. The proportionality factor then sets the
scale of distances in the non-perturbative QCD.

\section{Summary and Discussion}

The study reported here was  motivated by the recent experimental results
on the emission volume
size derived from the identical particle correlations measurements.
The information that this size is
seen to decrease as the mass of the hadron increases, is shown here
to be very valuable for the understanding of the non-perturbative
QCD mechanism.\\
\indent A good description of the $r$ dependence on the hadron mass
can already be obtained from the uncertainty principle. However
this requires the acceptance of two, not trivial and not obvious
assumptions, namely:
\vspace{-2mm}

\begin{enumerate}
\item\,\,\, $\Delta t$ does not depend on the mass value of the produced
particles. As is shown below,
this assumption is inconsistent with one of the models for
the non-perturbative description of the confinement forces, namely, with
the string approach.

\item\,\,\,$ \Delta E$ depends only on the kinetic energy of the produced
particle. It means that the interaction (potential energy) is rather
small.

\item\,\,\,$ \Delta r \,\sim \, r $ which is in the spirit with  
the uncertainty principle used here.
\end{enumerate}

\noindent
The second approach followed here is
based on the virial theorem and uses  the Local
Parton Hadron Duality \cite{DUAL}, which is the only argument
 why we can consider
the potential energy to be independent of the mass of the
produced particles.
It looks rather impressive that the phenomenological potential,
whose parameters were taken
from the characteristics of the hadron wave functions and decay
constants, is able to describe so successfully the experimental data.
This gives us therefore the hope that the experimental
data on the radius of the emission volume would be useful in any
discussion of the theoretical and/or phenomenological description  of the
non-perturbative QCD forces responsible for confinement of quarks and
gluons.\\
\indent The decrease of the emitter size with the mass of the produced hadrons
limits the models which can be applied to the
non-perturbative QCD. For example, consider the string approach to the
non-perturbative QCD, which is also the base of the Lund Model of
hadronisation described in detail in \cite{LUND}.
In this approach the formation time which we need to create a particle with
mass $m$, as well as the corresponding distance, are  
proportional to the mass of the produced particle, namely,\
$t(r)\,\propto\,m/\kappa$\  where $\kappa$ is the string
tension which appears in \eq{VT5}. The experimental data in Fig. 2   
shows in fact the opposite trend. Therefore, the string based models may have
difficulties in accommodating the data findings. On the other hand,
the perturbative QCD cascade plus the Local Parton Hadron Duality
\cite{DUAL} approach
leads to $r\,\propto \,1/m$ which reproduces qualitatively the
experimental data. These two examples clearly illustrate how these
measurements are able to select or reject theoretical proposals.
For
this reason more accurate measurements of\ $r_{\pi \pi}$,\, $r_{KK}$ and
\ $r_{\Lambda \Lambda}$ should be very valuable.\\
\indent We would like to stress  that the observed hierarchy of the sizes
 of the emission volume is providing us an additional information on the 
dynamics of interaction to that given by the  well established experimentally 
dependence of the effective source sizes on the produced particles
transverse mass, \mbox{$m_t .$} 
Experiments with  heavy-ions and hadron-hadron reactions  as well as LEP 
studies \cite{mt_exp}  show that \mbox{$r \,\sim\,1/\sqrt{m_t} .$}
In these experiments only Bose-Einstein correlations between identical
pions have been measured, where $r$ is  considered to be proportional to the
 production time \cite{csorgo}.
It should be stressed that this production time is defined as the invariant 
time elapsed between the creation of a $ q \bar{q}$ pair and the
formation of a pion.
 For large $p_t$ an hypothesis has been suggested 
that the production time is larger than the formation time.
This hypothesis 
is able to reproduce the experimental dependence of $r$ on $m_t$.
It is reasonable to assume  that the dependence of the size of the source
 on the mass of the produced particle is related to the formation time rather than
 to the production one. This is a reason why such experiments as 
discussed here  can give a new insight
 to the nonperturbative dynamics of hadron production.   

\section*{Acknowledgments}
We would like to thank W. Ochs for his valuable discussion and for
pointing out to us the different behaviour of the emission size
on the mass of the produced  hadron in the string and Local Parton Hadron Duality
models, which
he discussed  already in his 1986 CERN academic training
lectures. Our thanks also due to B. Andersson, A. Bia{\l}as, 
E. Gotsman, R. Lednicky, U. Maor, E.K.G. Sarkisyan and T. Sj\"{o}strand
for many valuable comments and discussions and for their 
keen interest in this work.



\end{document}